\documentclass[12pt,prd,preprint,showpacs]{revtex4}
\usepackage{amsmath}
\usepackage{amssymb}
\usepackage{graphicx}

\begin{document}

\raisebox{8mm}[0pt][0pt]{\hspace{9cm}\vbox{\hskip 1cm IFIC/04-30}}

\title{Enhanced solar anti-neutrino flux in random magnetic fields}
\author{O. G. Miranda$^1$} \email{Omar.Miranda@fis.cinvestav.mx}
\affiliation{$^1$Departamento de F\'{\i}sica, Centro de
  Investigaci{\'o}n y de Estudios Avanzados del IPN, Apdo. Postal 14-740 07000
  Mexico, DF, Mexico} \author{T. I. Rashba$^{2,3}$}
\email{rashba@izmiran.rssi.ru} \author{A. I. Rez$^{2,3}$}
\email{rez@izmiran.rssi.ru} \author{J.~W.~F. Valle$^2$}
\email{valle@ific.uv.es} \affiliation{$^2$Instituto de F\'{\i}sica
  Corpuscular -- C.S.I.C.,
  Universitat de Val{\`e}ncia \\
  Edificio Institutos, Apt.\ 22085, E--46071 Val{\`e}ncia, Spain}

\affiliation{$^3$ Institute of Terrestrial Magnetism,
    Ionosphere and Radio Wave Propagation of the Russian Academy of Sciences,
    142190, Troitsk, Moscow region, Russia}

\begin{abstract}
  
  We discuss the impact of the recent KamLAND constraint on the solar
  anti-neutrino flux on the analysis of solar neutrino data in the
  presence of Majorana neutrino transition magnetic moments and solar
  magnetic fields.  We consider different stationary solar magnetic
  field models, both regular and random, highlighting the strong
  enhancement in the anti-neutrino production rates that characterize
  turbulent solar magnetic field models.  Moreover, we show that for
  such magnetic fields inside the Sun, one can constrain the
  \emph{intrinsic} neutrino magnetic moment down to the level of
  $\mu_\nu \lesssim {\rm few} \times 10^{-12}\mu_B$ irrespective of
  details of the underlying turbulence model. This limit is more
  stringent than all current experimental sensitivities, and similar
  to the most stringent bounds obtained from stellar cooling. We
  also comment on the robustness of this limit and show that at most it
  might be weakened by one order of magnitude, under very unlikely
  circumstances.
   
\end{abstract}

\maketitle
\vskip 1.3cm

\section{Introduction}
\label{sec:introduction}
The KamLAND experiment has recently reached an improved sensitivity on
a possible electron anti-neutrino component in the solar
flux~\cite{Eguchi:2003gg}.  Their current limit corresponds to
$2.8\times10^{-2}\%$ of the solar boron $\nu_e$ flux, at the $90\%$
C.L., about 30 times better than the previous
Super-Kamiokande~\cite{Gando:2002ub} and the recent
SNO~\cite{Aharmim:2004uf} limits.
Solar anti-neutrinos constitute a characteristic signature of the
spin-flavor precession (SFP) mechanism when non-vanishing Majorana
neutrino transition magnetic moments~\cite{Schechter:1981hw} interact
with solar magnetic
fields~\cite{Schechter:1981hw,akhmedov:1988uk,lim:1988tk}.

The very first evidence of reactor anti-neutrino disappearance
published by the KamLAND collaboration~\cite{Eguchi:2002dm} has
already excluded SFP scenarios as solutions to the solar neutrino
problem~\cite{Barranco:2002te}. However this evidence still leaves
considerable room for sub-leading SFP effects in solar neutrino
physics.
It is the latest KamLAND limit on the solar electron anti-neutrino
flux~\cite{Eguchi:2003gg}, in combination with solar neutrino data,
including the recent SNO salt phase results~\cite{Ahmed:2003kj}, that
ultimately establishes the robustness of the simplest three-neutrino
oscillation description of the solar neutrino
data~\cite{Maltoni:2003da} showing how it is essentially stable even
in the presence of the SFP
mechanism~\cite{Miranda:2003yh}~\footnote{There is a loophole to this
  argument, namely the case of Dirac neutrinos. This is certainly a
  particular case of the SFP Hamiltonian~\cite{Schechter:1981hw} for
  which the anti-neutrino argument does not apply. However for this
  case the gauge theoretic expectations for the attainable magnitudes
  of the transition moment are significantly lower than those for
  Majorana neutrinos.  }.

Little is known about the detailed structure of solar magnetic fields
and several models have been previously used to analyse the SFP
conversion.  These solar magnetic field models make different
assumptions about the nature (regular or random) of solar magnetic
fields, their magnitude, location and typical
scales~\cite{Bykov:1998gv,Kutvitskii,Guzzo:1998sb,akhmedov:2002mf,Friedland:2002pg}.
According to the dynamo mechanism the solar magnetic field is
generated close to the bottom of the convective
zone~\cite{1983flma....3.....Z}.  Following this picture, we assume
that the field resides within the solar convective
zone~\cite{Bykov:1998gv,Kutvitskii,Guzzo:1998sb}. 
Moreover, in accordance with the present-day understanding of solar
magnetic field evolution, the large-scale magnetic field in the solar
convective zone is followed by a small-scale random component, whose
strength is expected to be comparable to or even larger than that of
the regular one.

Insofar as the SFP mechanism is concerned, the main difference between
random and regular magnetic field scenarios is that the former
generally give rise to an enhanced rate of anti-neutrino production,
up to two orders of magnitude when compared to the case of regular
fields of the same (average) amplitude.  This fact has been used in
Ref.~\cite{Miranda:2003yh} in order to obtain more stringent limits on
$\mu_{\nu}B$.  Moreover, assuming that random magnetic fields are of
turbulent origin we were able to extract a limit on the transition
magnetic moment itself ($\mu_{\nu}$).  We show that, under reasonable
assumptions, such bound is comparable to the best astrophysical limit
that follows from stellar cooling arguments~\cite{Raffelt:1990pj}.  An
alternative anaysis of the KamLAND data~\cite{Eguchi:2003gg} using the
"delta-correlated" model for the solar random magnetic field was done
in~\cite{Torrente-Lujan:2003cx}.

The main purpose of the present paper is to give a more comprehensive
description of the models and analysis already presented briefly in
Ref.~\cite{Miranda:2003yh}. First to elucidate the physics of the
enhanced anti-neutrino SFP conversion rates in random field models,
compared to regular magnetic field scenarios. We show how this follows
from the loss of coherence of the spin flavour evolution in a
fluctuating environment. Second, we show how in a solar
magnetohydrodynamics (MHD) turbulence model of Kolmogorov type one
can use the characteristic scaling in this theory in order to obtain a
limit on the intrinsic neutrino transition magnetic moment
$\mu_{\nu}$.

The paper is organized as follows.  In Sec.~\ref{sec:evolution} we
develop a perturbative approach to describe neutrino evolution in the
presence of convective--zone random magnetic fields, treating the
magnetic interactions as small correction to the oscillation evolution
Hamiltonian.  This provides a good approximation, fully justified in
view of recent KamLAND and solar neutrino data which support the MSW
LMA interpretation~\cite{Maltoni:2003da}.  We show that neutrinos
behave as a ``Fourier analyzer'' reading off only that spectral
harmonic of the two-point magnetic field correlation function whose
space period equals the neutrino oscillation length.
Our solar magnetic field model is discussed in
Sec.~\ref{sec:magfield}, first within the framework of the simplest
piece-constant correlation cell model with one effective correlation
scale $L_0$, and subsequently within a Kolmogorov--type turbulent
magnetic field picture.  Our discussion illuminates the difference
between SFP anti-neutrino production rates within random and regular
fields, as well as the physical meaning of the correlation cell model
parameters.  The results of our neutrino data analysis of the KamLAND
$\bar\nu_e$ limit and future perspectives are given in
Sec.~\ref{sec:new-sec4}. We also present in
Sec.~\ref{sec:crit-robustn-bound} a critique of the robustness of this
limit and show that at most it might be weakened by one order of
magnitude, under very unlikely circumstances.  Finally in
Sec.~\ref{sec:conclusions} we summarize our results.

\section{Neutrino Evolution}
\label{sec:evolution}

Here we adopt a simplified two-neutrino picture of neutrino evolution,
neglecting the angle $\theta_{13}$. As we show in the Appendix this is
a good approximation, in view of current neutrino data.  Solar
neutrino evolution in the presence of a magnetic field involves then
only the solar mixing angle where $\theta_{12} \equiv \theta_{sol}
\equiv \theta$ and is described by a four--dimensional
Hamiltonian~\cite{Schechter:1981hw,akhmedov:1988uk,lim:1988tk},
\begin{equation}
i\left(
\begin{array}{l}
         \dot{\nu}_{eL} \\
   \dot{\bar{\nu}}_{eR} \\
      \dot{\nu}_{a L} \\
\dot{\bar{\nu}}_{a R}
\end{array}
\right) = \left(
\begin{array}{cccc}
  V_e-c_2\delta  &                    0 & \quad         s_2\delta &   \mu_{\nu} b_+(t) \\
               0 &       -V_e-c_2\delta & \quad -\mu_{\nu} b_-(t) &          s_2\delta \\
       s_2\delta &    -\mu_{\nu} b_+(t) & \quad V_{a}+c_2\delta &                  0 \\
\mu_{\nu} b_-(t) &            s_2\delta & \quad                0 & -V_{a}+c_2\delta
\end{array}
\right) \left(
\begin{array}{c}
         \nu_{eL} \\
   \bar{\nu}_{eR} \\
      \nu_{a L} \\
\bar{\nu}_{a R}
\end{array}
\right)~,  \label{master}
\end{equation}
where $\nu_a=\nu_\mu\cos\theta_{23}-\nu_\tau\sin\theta_{23}$
($\theta_{23}\equiv\theta_{atm}$ is the atmospheric mixing angle);
$c_2 = \cos 2\theta$ and $s_2 = \sin 2\theta$; $\delta = \Delta
m^2/4E$ is assumed to be always positive. Note in this approximation
the Majorana neutrino transition magnetic moment element $\mu_\nu
\equiv \mu_{ea}$ describing transitions between neutrino flavour
states $\nu_e$ and $\nu_a$ coincides with the element $\mu_{12}$
characterizing transitions between mass eigenstates $\nu_1$ and
$\nu_2$; $V_e(t) = G_F\sqrt{2}(N_e(t) - N_n(t)/2)$ and $V_{a}(t) =
G_F\sqrt{2}(-N_n(t)/2)$ are the neutrino matter potentials for
$\nu_{eL}$ and $\nu_{a L}$ in the Sun, given by the number densities
of the electrons ($N_e(t)$) and neutrons ($N_n(t)$).  Finally,
$b_{\pm} = b_x \pm ib_y$ denote the magnetic field components which
are perpendicular to the neutrino trajectory.

Inside the radiative zone, where the magnetic field is neglected, the
evolution of the neutrinos reduces to that implied by the LMA MSW
oscillation hypothesis.  In order to get an approximate analytic
solution for Eq.~(\ref{master}) in the convective zone it is
convenient to work in the mass basis.  Defining the vectors
\begin{equation}
\nu_L = \left(
\begin{array}{c}
\nu_{1L} \\
\nu_{2L}
\end{array}
\right)
\qquad
\bar{\nu}_R = \left(
\begin{array}{c}
\bar{\nu}_{1R} \\
\bar{\nu}_{2R}
\end{array}
\right)~,
\end{equation}
we then express the evolution equation in a block form as
\begin{equation}
i\left(
\begin{array}{l}
         \dot{{\mathbf \nu}}_{L} \\
         \dot{\bar{{\mathbf \nu}}}_{R}
\end{array}\right) =  \left(
\begin{array}{cc}
H_{osc}           & H_{mag} \\
H_{mag}^{\dagger} & \bar{H}_{osc}
\end{array}
\right)  \left(
\begin{array}{c}
         \nu_{L} \\
    \bar{{\nu}}_{R} \\
\end{array}
\right)~,  \label{mmaster}
\end{equation}
where 
\begin{equation}
H_{osc} = \left(
\begin{array}{cc}
E_{1L} & -i\dot{\theta}_m \\
i\dot{\theta}_m &  E_{2L} \\
\end{array}
\right)  , \qquad
\bar{H}_{osc} = \left(\begin{array}{cc}
E_{1R} &  -i\dot{\bar\theta}_m \\
i\dot{\bar\theta}_m  & E_{2R}
\end{array}
\right)  
\end{equation}
and, 
\begin{equation}
H_{mag}=\left(
\begin{array}{ccc}
-\mu_{\nu} b_+ e^{i\Psi}\sin (\bar{\theta}_m - \theta_m )    &  \qquad &
 \mu_{\nu} b_+ e^{i\Psi}\cos (\bar\theta_m - \theta_m ) \\
-\mu_{\nu} b_+ e^{i\Psi}\cos (\bar\theta_m - \theta_m)       &  \qquad &
-\mu_{\nu} b_+ e^{i\Psi}\sin (\bar{\theta}_m - \theta_m)  
\end{array}
\right) 
\end{equation}
with
$\Psi= 1/2 \int{[V_e (t) + V_a(t)]dt}$. 
The neutrino eigen-energies $E_{iL}$ are defined by
\begin{equation}
E_{iL} = \mp \sqrt{(V-\delta c_2)^2+(\delta s_2)^2}\,,
\label{energy}
\end{equation}
where $V = (V_e - V_a)/2$ and the minus (plus) sign corresponds to
the energy state with $i=1$ ($2$). The solar neutrino mixing angle in
matter is given as
\begin{equation}
\tan 2\theta_m = -\frac{\delta s_2}{V-\delta c_2}\,. 
\label{theta}
\end{equation}
Similar expressions for the anti-neutrino eigen-energies and matter
mixing angle, $E_{iR}$ and $\bar{\theta}_m$, are easily obtained by
changing the sign of the matter potential, $V\to -V$, in
Eqs.~(\ref{energy}) and (\ref{theta}).

In the region of neutrino oscillation parameters indicated by current
neutrino data~\cite{Maltoni:2003da}, $\delta=\delta_{\rm LMA}$ and
$\theta=\theta_{\rm LMA}$, the matter effect in the convective zone
turns out to be rather small, since the ratio of the matter potential
$V$ to $\delta_{{\rm LMA}}$ is at most around $\sim 10^{-2}$, at the
bottom of the convective zone. This allows us to expand the neutrino
eigen-energies in powers of $V/\delta$
\begin{equation}
E_{iL} \simeq \mp \delta
\left[1-\frac{V}{\delta}c_2+\frac{V^2}{\delta^2}s_2^2+
O\left(\frac{V^3}{\delta^2}\right) \right]
\label{energy1}
\end{equation}
and similarly the neutrino mixing angle in matter may be expressed as
\begin{equation}
\theta_m \simeq \theta + \frac{V}{\delta}s_2 -
\frac{V^2}{\delta^2}(1+s_2^2) \tan 2\theta
+O\left(\frac{V^3}{\delta^2}\right)~.
\label{theta1}
\end{equation}

An analogous estimate gives an upper bound on the mixing angle
derivative, $|\dot{\theta}_m| /\delta \lesssim 2\times 10^{-5}$ and
\begin{equation}
\left | \frac{\dot{\theta}_m}{\mu_\nu b_\perp} \right | \lesssim 5\times
10^{-3} \left(\frac{10^{-11}\mu_B}{\mu_\nu}\right) \left(\frac{100
{\rm kG}}{b_\perp}\right)\,.
\end{equation}
with similar approximations valid for anti-neutrinos.

Therefore we can safely neglect all powers of $V/\delta$, along with
$\dot{\theta}_m$ and $\dot{\bar{\theta}}_m$. In this approximation the
$(4\times 4)$ evolution equation decouples into two $(2\times
2)$~equations
\begin{equation}
i\left(
\begin{array}{l}
\dot{\nu}_{1L} \\
\dot{\bar{\nu}}_{2R}
\end{array}
\right) = \left(
\begin{array}{cc}
-\delta &   \mu b_+(t) e^{i\Psi} \\
\mu b_-(t)e^{-i\Psi} &     \delta
\end{array}
\right) \left(
\begin{array}{c}
\nu_{1L} \\
\bar{\nu}_{2R}
\end{array}
\right)~, \label{msseq} 
\end{equation}
\begin{equation}
i\left(
\begin{array}{l}
\dot{\nu}_{2L} \\
\dot{\bar{\nu}}_{1R} 
\end{array}
\right) = \left(
\begin{array}{cccc}
      \delta & - \mu b_+(t) e^{i\Psi}\\
- \mu b_-(t)e^{-i\Psi} &    - \delta  \\
\end{array}
\right) \left(
\begin{array}{c}
\nu_{2L} \\
\bar{\nu}_{1R} 
\end{array}
\right)~,   \label{masseq}
\end{equation}
describing spin-flavour precession of the two mass-eigenstate pairs.
In the absence of magnetic fields the neutrino eigenstates propagate
independently across the convective zone. When the magnetic field is
switched on one obtains a mixing, characteristic of the spin flavour
precession mechanism~\cite{Schechter:1981hw}.  This turns out to be
rather small, a fact that greatly simplifies the problem of solving
the evolution equation.  For this we consider the parameter $\kappa$
defined as
\begin{equation}
\kappa = \frac{\mu_\nu^2 b_\perp^2}{\delta^2}
= 2.5 \times 10^{-5} \left(\frac{\mu_\nu}{10^{-11}\mu_B}\right)^2 
\left(\frac{b_{\perp max}}{100 {\rm kG}}\right)^2 
\left(\frac{7\times 10^{-5} {\rm eV^2}}{\Delta m^2}\right)^2 
\left(\frac{E}{10{\rm MeV}}\right)^2~.
\label{kappa}
\end{equation}
The smallness of $\kappa$ allows us to expand the neutrino survival
probabilities at the surface of the Sun in powers of $\kappa$, to any
given desired accuracy in perturbation theory.
The matter effect~\cite{akhmedov:1988uk,lim:1988tk} is also rather
small, and is fully encoded by the phase $\Psi$, so that when $\Psi
\to 0$ one recovers the vacuum result of Ref.~\cite{Schechter:1981hw}.

From Eqs.~(\ref{msseq}) and (\ref{masseq}) we find that to leading
order in $\kappa$ the neutrino mass eigenstate probabilities at the
surface of the Sun can be written in the form
\begin{equation}
\label{mprob1}
|\nu_{1L}|^2_{R_\odot} = P_1(1-\eta), \qquad  
|\nu_{2R}|^2_{R_\odot} = P_1 \eta~,
\end{equation}
\begin{equation}
|\nu_{2L}|^2_{R_\odot} = P_2(1-\eta), \qquad  
|\nu_{1R}|^2_{R_\odot} = P_2 \eta~,
\label{EqnAtSunSurface}
\end{equation}
(note that unitarity is fulfilled), where the key small parameter
$\eta$ is given by
\begin{equation}
\eta = \frac{\mu^2}{2}\int^L_0 dt_1 \int^L_0 dt_2 [b_+(t_1)b_-(t_2)
e^{-2i\delta(t_1-t_2)} + c. c.]~.
\label{eta}
\end{equation}
Here $L$ is the width of the convective zone and $P_i
=|\nu_{iL}(r=0.7R_\odot)|^2$ denote the probabilities that solar
neutrinos reach the bottom of the convective zone in a given mass
state, calculated numerically in the LMA-MSW oscillation picture,
similar to the method used in \cite{Maltoni:2003da}.

The probabilities for neutrinos to be detected in flavour states
$\alpha=e,\mu,\bar{e},\bar{\mu}$, $P_{e \alpha}$, are then given by
\begin{eqnarray}
P_{ee}&=&\left[P_1P_{1e}+P_2P_{2e}+2\sqrt{P_1P_2P_{1e}P_{2e}}\cos\xi_1\right](1-\eta),\\
\label{Pee}
P_{e\mu}&=&\left[P_1P_{1\mu}+P_2P_{2\mu}-2\sqrt{P_1P_2P_{1\mu}P_{2\mu}}\cos\xi_1\right](1-\eta),\\
\label{Pemu}
P_{e\bar{e}}&=&\left[P_{2}P_{\bar{1}\bar{e}}+P_{1}P_{\bar{2}\bar{e}}-2\sqrt{P_{2}P_{1}P_{\bar{1}\bar{e}}P_{\bar{2}\bar{e}}}\cos\xi_2\right]\eta,\\
\label{Peae}
P_{e\bar{\mu}}&=&\left[P_{2}P_{\bar{1}\bar{\mu}}+P_{1}P_{\bar{2}\bar{\mu}}+2\sqrt{P_{2}P_{1}P_{\bar{1}\bar{\mu}}P_{\bar{2}\bar{\mu}}}\cos\xi_2\right]\eta .
\label{Peamu} \label{probabilities}
\end{eqnarray}
Here $P_{i \alpha}$ is the $\nu_i \to \nu_\alpha$ conversion
probability from the surface of the Earth to the detector.  The phases
$\xi_k$ ($k=1,2$) characterize the evolution in vacuum from the Sun to
the Earth, and are given by
\begin{equation}
\xi_k=\frac{\Delta m^2 (D+L-R_\odot)}{2E}+\phi_k\, ,
\label{int}
\end{equation}
where $D$ is the Sun-Earth distance, $\phi_k$ contain the phases due
to propagation in the Sun up to the bottom of the convective zone and
in the Earth. We have checked that $\phi_k$ can be safely neglected for
our purposes.

In summary, neutrino evolution can be understood as follows:
\begin{itemize}
\item neutrinos, generated in the solar core, undergo LMA MSW
  conversion and enter the convective zone as a coherent mixture of
  $\nu_{eL}$ and $\nu_{a L}$.  By numerically solving the
  corresponding $(2\times 2)$ MSW evolution problem we obtain the
  amplitudes $\nu_{eL}(0)$ and $\nu_{a L}(0)$ at the bottom of the
  convective zone, which are then used as initial values for neutrino
  propagation across the convective zone.
\item neutrino evolution within the convective zone is considered as
  approximately vacuum oscillations modulated by a small spin-flavour
  conversion~\cite{Schechter:1981hw}. This is treated in leading order
  in the small expansion parameter $\kappa$, Eq.~(\ref{kappa}).
\item the neutrino survival and conversion probabilities, defined at
  the solar surface, are then evolved to the detector taking into
  account regeneration in the Earth.
\end{itemize}
This way neutrino and anti-neutrino yields are determined and used in
the analysis of solar neutrino and KamLAND data.

\section{Magnetic Field Model}
\label{sec:magfield}
The main goal of this section is to calculate the parameter $\eta$ in
Eq.~(\ref{eta}) characterizing the solar anti-neutrino production rate
to first order in the small expansion parameter $\kappa$ in
Eq.~(\ref{kappa}) describing neutrino propagation in the framework of
different solar magnetic field models.

Different approaches have been used in order to describe neutrino
propagation in the fluctuating matter density or random magnetic field
environments.
Since the origin of the media fluctuations is unknown, we stick to the
simplest ``white'' noise description, which involves just two
parameters: the typical correlation scale and its characteristic
amplitude.

The ``delta-correlated'' model described by
\[
\left\langle b_\perp(r_1)b_\perp(r_2)\right\rangle = 
\bar{b_\perp^2}\left(\frac{r_1+r_2}{2}\right)
L_0\delta (r_1-r_2)
\]
has been used both for the case of matter density
fluctuations~\cite{Loreti:1994ry,Nunokawa:1996qu,Balantekin:2003qm,Guzzo:2003xk},
as well as random magnetic
fields~\cite{Nicolaidis:1991da,Bykov:1998gv,Semikoz:1998ef,Torrente-Lujan:2003cx}.
This choice of correlator has the virtue that it allows for analytical
solutions to the neutrino evolution equations.
It suffers from an important limitation, namely that the allowed
spatial scales, $L_0$, of fluctuations should be less than typical
neutrino oscillation lengths $\lambda_{osc}=4\pi E / \Delta m^2$.
However, it so happens that the strongest conversion effect takes
place when these two parameters are comparable~\cite{Burgess:1997mz}.

Because of this one needs to go at least one step further and consider
the so called ``piece-constant'' model. For the case of matter density
fluctuations this has been used in
Refs.~\cite{Burgess:1997mz,Burgess:2002we,Burgess:2003su} and for
random magnetic fields it was applied in~\cite{Bykov:1998gv}. 

In what follows (Sec.~\ref{sec:simpl-rand-field}) we adopt this
piece-constant model, whose correlator is described by,
\begin{eqnarray}
\left\langle b_\perp(r_1)b_\perp(r_2) \right\rangle = 
\bar{b_\perp^2}\left(\frac{r_1+r_2}{2}\right)\phantom{0}, 
&{\rm if}& |r_1-r_2| \leq L_0  \nonumber\\
\left\langle b_\perp(r_1)b_\perp(r_2)\right\rangle = 
0\phantom{\bar{b_\perp^2}\left(\frac{r_1+r_2}{2}\right)}, &{\rm if}& 
|r_1-r_2| > L_0. 
\nonumber
\end{eqnarray}
In other words, the fluctuation correlation function is modeled as a
step-function.  We refer to this as ``simplest random magnetic field
model''.

Note that in these models the typical correlation scale is in general
a free unknown parameter. One may go a step further if one has more
information about the nature of the assumed fluctuations. For
instance, a detailed analysis of how density fluctuations, in the form
of helioseismic waves, can affect the MSW neutrino oscillations was
introduced in Ref.~\cite{Bamert:1998jj}. In the discussion we give in
Sec.~\ref{sec:magn-hydr-turb} we use the solar MHD turbulence model in
order to describe the nature of the solar random magnetic fields.
These are described by a magnetic field correlation tensor
\[
\left\langle b_i(r_1)b_j(r_2)\right\rangle = M_{ij}(r_1,r_2)\,,
\]
whose specific form will be given below (Eq.~(\ref{correl})).  As a
result the separate dependence on the correlation scale and the
amplitude of fluctuations is replaced by a dependence on a specific
combination of these parameters, $\varepsilon$ in
Eq.~(\ref{FetaRenormalized}). This has the advantage of expressing the
final neutrino fluxes in terms of a single effective parameter which
varies over a relatively narrow range.

\subsection{Simplest random field model}
\label{sec:simpl-rand-field}

As already mentioned, current views on solar magnetic field evolution
suggest that the mean large-scale magnetic field is followed by a
comparable small-scale random magnetic field component. Such random
small-scale magnetic field is not directly traced by sunspots or other
tracers of solar activity. Dynamically, this field propagates through
the convective zone and photosphere drastically decreasing in
strength.  While we lack a direct reliable observational estimate of
its amplitude, one finds that the ratio of the random to regular
magnetic field amplitudes may be as large as 50-100, as it is not
clear at what stage the dynamo mechanism saturates. This issue has
been very actively discussed in the literature (see,
e.g.,~\cite{1983flma....3.....Z,Subramanian:1999mr} and references
therein).

The simplest random convective--zone solar magnetic field model is
obtained by imagining that the convective zone consists of a set of
correlation cells of volume $L_0^3$ where the random magnetic field is
assumed uniform, fields in adjacent cells being uncorrelated.  In this
picture we treat the small-scale random magnetic fields in terms of a
single effective scale $L_0$ characterizing the size of the
correlation cells.
We also assume that within each cell different magnetic field
components transversal to the neutrino trajectory are independent
random variables with zero mean value (for a more detailed discussion
see, for example, Ref.~\cite{Bykov:1998gv}).  That is, a given
realization of the random magnetic field along the neutrino path is a
stationary random process described by Gaussian statistics.  It is
also assumed that in order to satisfy divergence-less condition,
$\nabla \cdot \mathbf{B} = 0$, the magnetic field strength changes
smoothly at the boundaries between adjacent cells. This simplified
model seems reasonable since we do not expect a strong influence of
details of the random magnetic field structure near the layers
separating adjacent domain cells.

In order to estimate $\eta$ in Eq.~(\ref{eta}) we divide the range of
integration into a set of equal intervals of correlation length $L_0$
and average over random magnetic fields in each correlation cell,
\begin{equation}
\left\langle \eta \right\rangle
 = \frac{1}{2} \mu^2
  \sum^N_{k=1} \sum^N_{l=1}
  \int^{kL_0}_{(k-1)L_0} dt_1
  \int^{lL_0}_{(l-1)L_0} dt_2
\left\langle 
b_+(t_1)b_-(t_2) e^{-2i\delta(t_1-t_2)} + c.c.
\right\rangle\,,
  \label{fin_differ}
\end{equation}
this results in the final equation
\begin{equation}
\left\langle 
\eta\right\rangle
 = \left(\sum^N_{n=1} \mu^2 \bar{b^2}_{\perp n}\right)
\frac{\sin^2\left(\delta\cdot L_0\right)}{\delta^2}\,,
\label{aveta}
\end{equation}
where $\bar{b^2}_{\perp n}$ is the averaged square of the random
magnetic field in the $n$-th correlation cell. We assume that the root
mean square ({\it rms}) field varies smoothly along the neutrino
trajectory.  One therefore clearly sees the cumulative effect
characterizing neutrino propagation in random magnetic fields, implied
by the sum in Eq.~(\ref{aveta}).  For the simple case where all {\it rms}
field amplitudes in different cells are equal to some common magnetic
field value the above result gets proportional to the number of
correlation cells traversed by the neutrino, $N = L/L_0$.

In contrast, for regular magnetic fields the situation is different.
The neutrino to anti-neutrino conversion probability after traversing
the convective zone with a constant regular magnetic field of the same
amplitude is proportional to
\begin{equation}
\label{regfield}
\eta = \frac{\mu_{\nu}^2 b_{\perp}^2}{\delta^2 + \mu_{\nu}^2 b_{\perp}^2}
\sin^2 \left(\sqrt{\delta^2 + \mu_{\nu}^2 b_{\perp}^2} L\right) =
 \frac{\mu_{\nu}^2 b_{\perp}^2}{\delta^2}
\sin^2  (\delta \cdot L) + O\left(( \frac{\mu_{\nu}^2 b_{\perp}^2}{\delta^2})^2 \right)
\approx  \frac{\mu_{\nu}^2 b_{\perp}^2}{2\delta^2}~,
\end{equation}
that is, what would be expected after passing only one cell.  

Therefore in the random magnetic field case one obtains a sizable
enhancement of the neutrino conversion probability as compared with
the case of a constant magnetic field of the same amplitude.  This
enhancement is explained by the fact that the random nature of the
magnetic field destroys the coherence in the neutrino evolution.
Therefore, instead of adding amplitudes one have to add
probabilities~\cite{rez04}.

Finally, in order to take account of the shape of the {\it rms} random
field profile we introduce a factor
\begin{equation}
S^2=\frac{1}{N}\sum\limits_{n=1}^{N}\frac{\bar{b_n^2}}{b_{\perp max}^2},
\label{S}
\end{equation}
which is $S=1$ for constant {\it rms} field and of the order of unity for
other sufficiently wide spatial profiles, e.g.  $S\approx0.579$ for
``smooth'' profile~\cite{Bykov:1998gv}, $S\approx0.577$ for triangle
profile~\cite{Guzzo:1998sb}, $S\approx0.782$ for Kutvitsky-Solov'ev
profile~\cite{Miranda:2000bi}.
An alternative estimate of typical shape factors comes from the
assumption that the magnetic energy density globally follows the
approximate solar density profile with the scale height $H=0.1
R_{\odot}$, leading to $S \approx \sqrt{H/L} \approx
1/\sqrt3\approx0.57$.
 
Using the above definition of the shape factor one can rewrite
Eq.~(\ref{aveta}) as 
\begin{equation}
\left\langle 
\eta\right\rangle
 = \frac{ \mu^2 b^2_{\perp max}}{\delta^2}S^2~\frac{L}{L_0}~
\sin^2 (\delta \cdot L_0)\,,
\label{Saveta}
\end{equation}
where $ b^2_{\perp max}$ is the maximal value of the transverse
magnetic field.  This shows that different random magnetic field
models can be effectively characterized by two parameters,
$b^2_{\perp max} S^2$ and $L_0$.

\subsection{Magnetohydrodynamic turbulence models}
\label{sec:magn-hydr-turb}

A well motivated class of magnetic field models can be considered for
which the number of parameters can be further reduced by eliminating
reference to the effective scale $L_0$.  The relevant average
$\left\langle (b_+(t_1)b_-(t_2)\right\rangle$ of the transverse
components of the magnetic fields can be characterized by introducing
the two-point magnetic field correlation tensor as $M_{ij} =
\left\langle b_i(\mathbf{r}_1)b_j(\mathbf{r}_2)\right\rangle$.

For the case of isotropic, homogeneous and non-helical random magnetic
fields $M_{ij}$ is separated as
\begin{equation}
M_{ij}~=~M_N\left(\delta_{ij}-\frac{r_i r_j }{r^2}\right)+
M_L~\frac{r_i r_j }{r^2}~.
\label{correl}   
\end{equation}
where $\mathbf{r} = \mathbf{r}_1 - \mathbf{r}_2$. The longitudinal
($M_L$) and transverse ($M_N$) correlation functions depend only on
the separation distance between the two points, $r=|\mathbf{r}_1 -
\mathbf{r}_2|$.  Given that $\mathbf{\nabla}\cdot \mathbf{B}=0$, one
can write
\begin{equation}
\label{link}
M_N(r) = \frac{1}{2r} \frac{\partial}{\partial r} (r^2 M_L (r))~.
\end{equation}

One can generalize the above definition of the correlation function in
Eq.~(\ref{correl}) so as to cover the case of media which are
isotropic and homogeneous only {\it locally}.  This can be done by
expressing the correlator in factorized form as~\cite{MoninYaglom}
\begin{equation}
M_N (\mathbf{r_1},\mathbf{r_2}) = F\left( \frac{\mathbf{r_1}+\mathbf{r_2}}{2}\right) K_N (|\mathbf{r_1}-\mathbf{r_2}|)~, 
\label{splitted}
\end{equation}
where the random magnetic field profile factor $F$ depends on the
center-of-mass position of the two points $\mathbf{r_1}$ and
$\mathbf{r_2}$ and the local correlator $K_N(r)$ coincides with the
one characterizing the isotropic and homogeneous case, $M_N (r)$,
given in~Eq.~(\ref{correl}).

The above form provides a reasonable description for solar magnetic
field profiles.  Indeed, typical solar {\it rms} field profiles ($F$) vary
on scales of the order of the density scale height $H \approx
70000~\rm{km}$. On the other hand $K_N$ involves averaging on much
smaller scales $\delta ^{-1}$ related to the neutrino oscillation
length, about hundred kilometers for the LMA-MSW case.

Substituting Eq.~(\ref{correl}) into Eq.~(\ref{eta}) and restricting
the coordinates to the neutrino path, we get
\begin{equation}
\left\langle 
\eta\right\rangle
 = \mu^2\int^L_0 dz_1 \int^L_0 dz_2 M_N(z_1,z_2)
\cos [2\delta (z_1-z_2)] ~,
\label{Aveta}
\end{equation}
where $z_1$ and $z_2$ are two points on the neutrino trajectory.

Substituting Eq.~(\ref{splitted}) into Eq.~(\ref{Aveta}), changing
variables and rearranging we get
\begin{equation}
\label{feta}
\left\langle 
\eta\right\rangle
 = 4 \mu ^2 S^2 L \int^\infty_0 d\xi K_N(\xi) \cos (2\delta
\cdot \xi) ~,
\end{equation}
where the integration is extended to infinity because of $\delta \cdot
H \gg 1$ and the shape factor $S^2$ is defined as a continuous
analogue of Eq.~(\ref{S})
\begin{equation}
S^2=\frac{1}{L}\int\limits_{0}^{L}\frac{\bar{b_\perp^2}(z)}{b_{\perp max}^2}
dz\,.
\label{Scont}
\end{equation}

In order to estimate $K_N(\xi)$ we assume that magnetic field
evolution in the solar convective zone is due to the highly developed
steady-state MHD turbulence treated within the Kolmogorov scaling
theory~\cite{Kolmogorov:1941,Kolmogorov:1991,1983flma....3.....Z}. In
other words, for large magnetic Reynolds number, $R_m \sim
10^8$~\cite{1983flma....3.....Z}, the solar MHD turbulence is pumped
by the largest eddy motion that follows from the interplay between
convection and differential rotation of the Sun.  The size of the
largest eddies, $L_{{\rm max}}$, may be associated with the solar
granule size of the order of $1000~{\rm km}$.  Dynamo enhancement
subsequently results in the direct cascade of the energy of MHD
fluctuations to smaller
scales~%
\footnote{Note that small-scale MHD fluctuations may in turn produce
  regular large-scale fields through an inverse cascade mechanism. By
  large-scale here we mean scales much larger than the outer scale
  $L_{{\rm max}}$ ({\it i. e.}, the largest eddy scale) of the turbulence.}.
The smallest scale at which turbulent motion starts to decay
transferring energy into heat is the dissipative scale defined through
$l_{{\rm diss}}=L_{{\rm max}}R_m^{-3/4} \approx 1~{\rm m}$~. 

Within the inertial range, $l_{{\rm diss}}<l<L_{{\rm max}}$, the
similarity arguments of the Kolmogorov theory require the turbulent
hydrodynamic (HD) kinetic energy spectrum to scale as $E_{{\rm
    HD}}\sim k^{-5/3}$, where $k\sim 1/l$ is the wave number of the
eddies of size
$l$~\cite{Kolmogorov:1941,Kolmogorov:1991,1983flma....3.....Z}.
Similar qualitative arguments applied to the MHD case imply that the
corresponding Iroshnikov-Kraichnan scaling law can be taken as
$E_{{\rm MHD}}\sim k^{-3/2}$~\cite{1983flma....3.....Z}.  However
recent theoretical results and numerical simulations suggest that the
simpler HD Kolmogorov spectrum can actually be used even for the MHD
case~\cite{Goldreich,Cho}.  For our purposes the difference in the
power-law exponents turns out to be not specially important so that,
we take for definiteness the Kolmogorov value
$p=5/3$.  Generalization to arbitrary $p$'s is straightforward. See
details below.

Let us model the longitudinal correlation function $K_L(r)$ as in
hydrodynamics~\cite{MoninYaglom}
\begin{equation}
\label{KL}
K_L(l) = \frac{\bar{b^2}}{3}\frac{2^{2/3}}{\Gamma (1/3)}
\left(\frac{l}{L_{\rm max}}\right)^{1/3} K_{1/3}\left(\frac{l}{L_{\rm
max}}\right)~,
\end{equation}
where $\Gamma (x)$ is the gamma-function, $K_{\mu}(x)$ is the McDonald
function of index $\mu$ and $\bar{b^2}$ is the squared {\it rms} magnetic
field on scale $L_{\rm max}$. The model function correctly reproduces
required asymptotics of the Kolmogorov theory:
\begin{itemize}
\item[(i)] 
$K_L (0) = 
\left\langle 
b_z^2(z)\right\rangle
 = \bar{b^2}/3$~, 
\item[(ii)] 
$K_L (0) - K_L (l) \sim {l}^{2/3}$ for $l \ll L_{\rm max}$~.
\end{itemize}

Using Eq.~(\ref{link}) and Eq.~(\ref{KL}) and taking into account that
$2\delta \cdot L_{{\rm max}}\gg 1$ we finally perform the integration
in Eq.~(\ref{feta}) and obtain \footnote{Note that the exponential
  tail of Eq.(\ref{KL}) for $r \ge L_{{\rm max}}$ does not contribute
  to Eq.(\ref{Feta}) being suppressed by the fast oscillating
  integrand of Eq.~(\ref{feta}).}
\begin{equation}
\label{Feta}
\left\langle 
\eta\right\rangle
 \simeq \frac{\sqrt{\pi}}{3}\frac{2^{1/3}\Gamma (5/6)}{\Gamma
(1/3)} \frac{{\mu}^2 \bar{b^2} S^2 L L_{{\rm max}}}{(\delta\cdot L_{{\rm
max}})^{5/3}} \simeq 
0.3\frac{{\mu}^2 \bar{b^2} S^2
}{{\delta}^2} \frac{\delta\cdot L}{(\delta\cdot L_{\rm{max}})^{2/3}}~.
\end{equation}
which leads, in normalized units, to
\begin{equation}
\label{FetaRenormalized}\
\left\langle 
\eta\right\rangle
 \simeq 3\times 10^{-3} 
\mu_{11}^2 \varepsilon^2 S^2 
\left(\frac{7\times 10^{-5} {\rm eV}^2}{\Delta m^2} \right)^{5/3}
\left(\frac{E}{10 {\rm MeV}}\right)^{5/3}\,,
\end{equation}
where $\mu_{11}$ is the magnetic moment in units of $10^{-11}\mu_B$,
and the ratio $\varepsilon= (b/100~{\rm kG})/(L_{{\rm max}}/1000~{\rm
km})^{1/3}$.

The ratio $\varepsilon$ is not known precisely, but one may estimate
it assuming equipartition between kinetic energy of hydrodynamic
fluctuations and the {\it rms} magnetic energy at the largest (most
energetic) scale $L_{max}$~\cite{1983flma....3.....Z}
\begin{equation}
\frac{\rho \bar{v^2}}{2}\approx \frac{\bar{b}^2}{8\pi}
\end{equation}
Taking $v\sim 3\times 10^4 {\rm cm}/{\rm s}$ and $\rho \sim 1~{\rm
  g/cm}^3$ ~\cite{1991sun..book.....S} we obtain typical amplitude for
magnetic field fluctuations $b\sim100$~kG at the scale
$L_{max}=1000$~km.  Assuming that $b\approx 50-100$~kG and that the
shape factor $S$ lies in the range between 0.5 and 1 we estimate that
the product $\varepsilon S$ may vary in the interval $0.25<
\varepsilon S <1$.

Before concluding this section we note that both types of magnetic
field models lead to qualitatively similar results for the neutrino
conversion parameter $
\left\langle 
\eta\right\rangle
$ when the magnetic field correlation
scale is chosen to coincide with neutrino oscillation length
$\lambda_{osc}$. Indeed, in accordance with the Kolmogorov scaling law
$\bar{b^2}/(\delta \cdot L_{\rm{max}})^{2/3}\simeq \lambda_{\rm
  osc}^{2/3}\cdot \bar{b^2}/L_{\rm{max}}^{2/3}= \bar{b}_{\lambda_{\rm
    osc}}^2$ is the squared {\it rms} field at the scale $\lambda_{\rm
  osc}$.  This implies that Eq.~(\ref{Saveta}) goes into
Eq.~(\ref{Feta}) when making the replacement $\delta \cdot L\sim
L/\lambda_{\rm osc}$. This happens because in the context of the
turbulent magnetic field model neutrinos effectively feel only one
scale, namely their oscillation length.


\section{Limits on neutrino magnetic moments}
\label{sec:new-sec4}

In this section we analyze the limit on electron anti-neutrino flux
published by KamLAND~\cite{Eguchi:2003gg}.  In Ref.~\cite{Miranda:2003yh} we
showed how this limit makes the determination of neutrino oscillation
parameters, $\Delta m^2_{sol}$ and $\theta_{sol}$, extremely robust
against possible existence of spin-flavor conversions. This can be
used in order to determine the allowed regions of solar neutrino
oscillation parameters independently of the magnetic field and
magnetic moment parameters.

\begin{figure}[htbp]
  \centering
\includegraphics[width=.7\columnwidth]{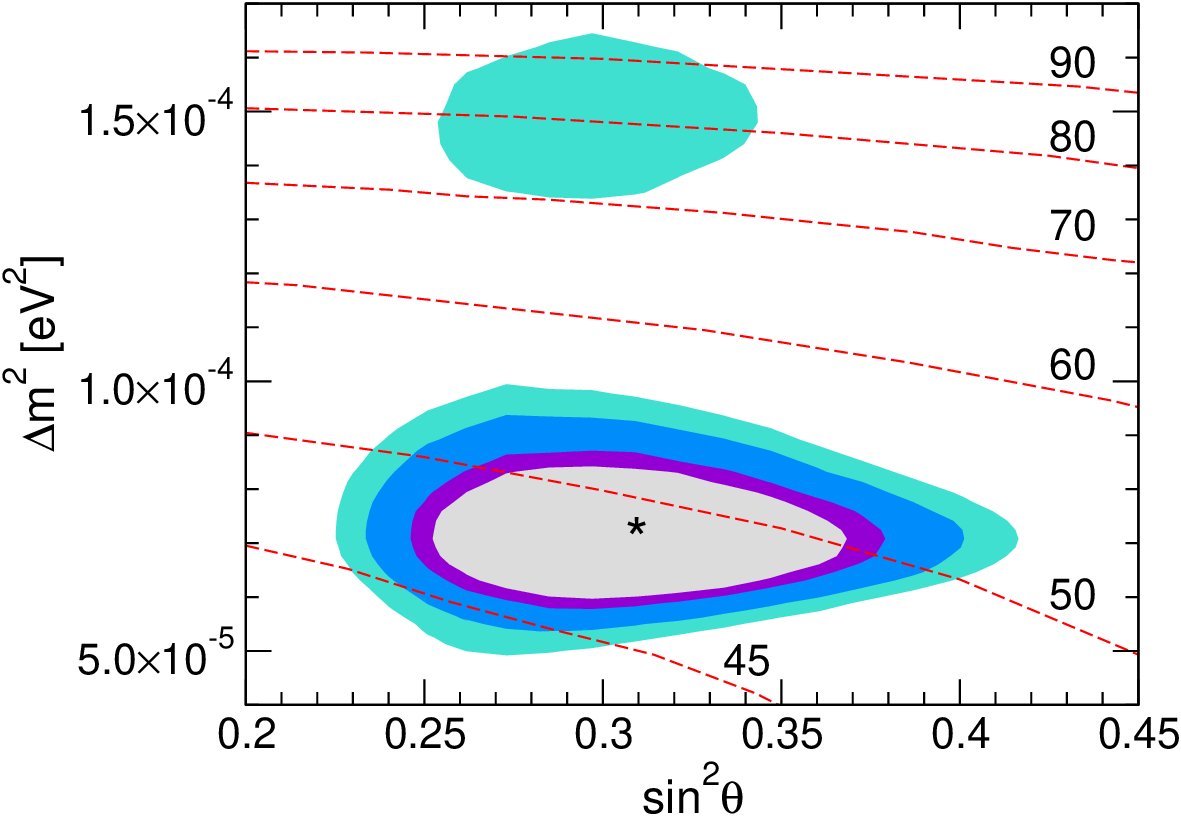}
\caption{\label{fig:iso-random} 
  Allowed regions for $\Delta m^2_{sol}$ and $\theta_{sol}$ from solar
  and KamLAND neutrino experiments at 90\%, 95\%, 99\% and 99.73\%
  C.L.  The curves are electron anti-neutrino iso-flux contours for
  $L_0=100$~km, plotted for different $\mu_{11}S b_\perp$ values
  indicated in kG, and fixing the anti-neutrino production at the
  current KamLAND limit. For values above 55 kG or so the high LMA
  region is ruled out.}
\end{figure}

Using the standard $\chi^2$ procedure (see~\cite{Barranco:2002te} and
references therein) and taking into account full set of solar as well
as KamLAND reactor neutrino data we have re-determined the allowed
regions of solar neutrino oscillation parameters, $\Delta m^2_{sol}$
and $\theta_{sol}$, within the recent version of Standard Solar Model
(BP04)~\cite{Bahcall:2004fg}.  The results are presented as shaded
regions in Fig.~\ref{fig:iso-random}.

Let us first consider the simple random magnetic field model described
in Sec.~\ref{sec:simpl-rand-field}. In this case neutrino conversion
probabilities depend both on the oscillation parameters, $\Delta
m^2_{sol}$ and $\theta_{sol}$, as well as the parameters $\mu^2_\nu
b^2_{\perp_{max}}S^2$ and $L_0$ describing the random magnetic field
model.  Using Eqs.~(\ref{Pemu}) and (\ref{aveta}) we have calculated
the predicted electron anti-neutrino flux.  In
Fig.~\ref{fig:iso-random} we show, for a fix value of $L_0=100$~km and
$S^2=1$, the curves that correspond to an electron anti-neutrino yield
of $2.8\times 10^{-4}\phi_B$. It is clear that a better determination
of the solar mixing angle by future experiments will not substantially
improve the limits on the parameters $\mu^2_\nu b^2_{\perp_{max}}S^2$
and $L_0$ which are mainly restricted by the solar anti-neutrino flux
limit.  In contrast note that an improved determination of the solar
mass splitting at KamLAND will play an important role in pinning down
the magnetic field parameters.

\begin{figure}[htbp]
  \centering
\includegraphics[width=.7\columnwidth]{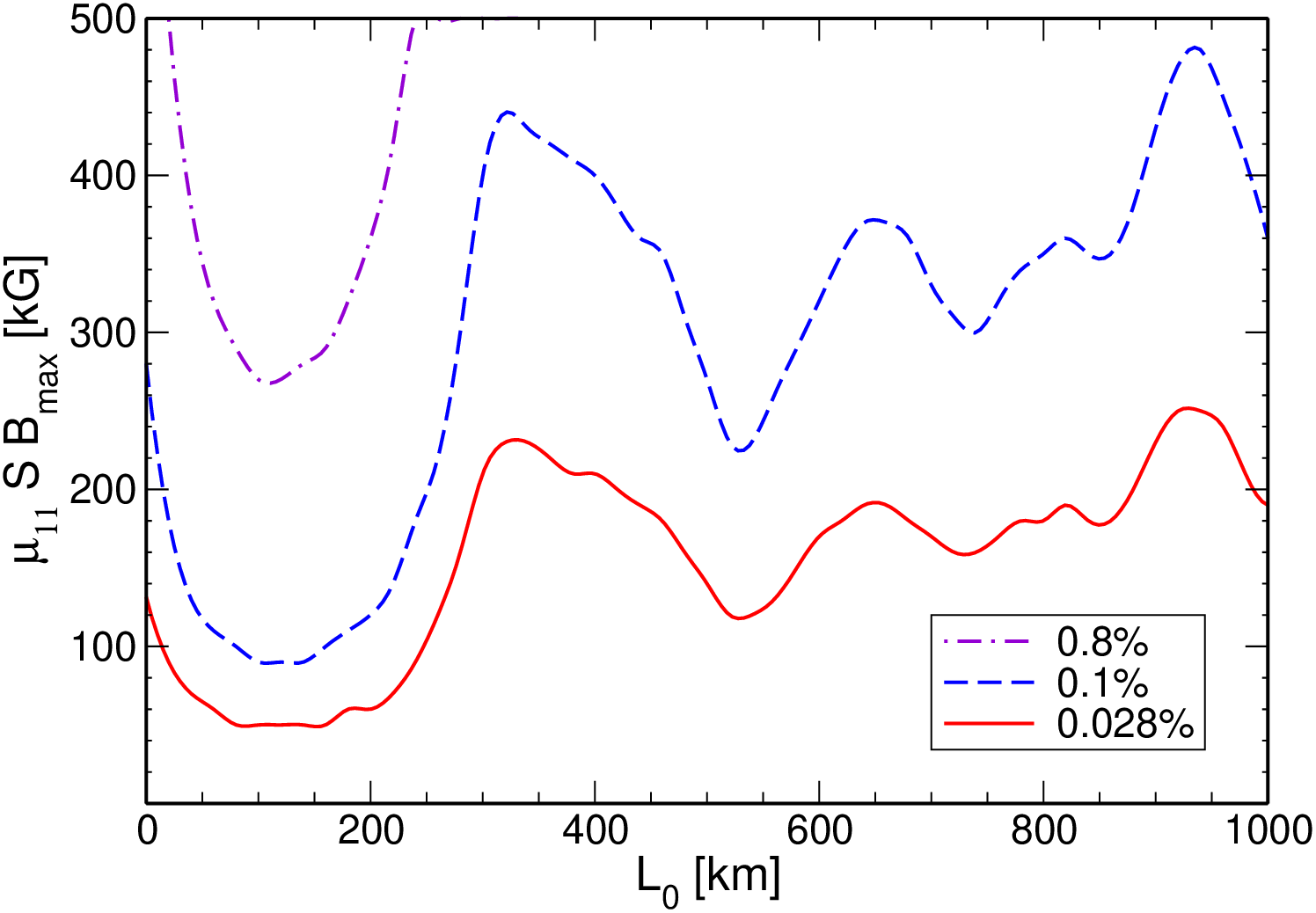}
\caption{\label{fig:random} 
  Upper bound on $\mu_{11} S b_\perp$ for a random field with spatial
  scale $L_0$ allowed by the 90\% C. L. region of $\Delta m^2_{sol}$
  and $\theta_{sol}$. The solid line shows the allowed values coming
  from the current KamLAND limit on $\phi_{\bar\nu_e}$.  For
  comparison we also give the limits that would correspond to
  $\phi_{\bar\nu_e}=0.1 \%$ and $\phi_{\bar\nu_e}=0.8$ \% of
  $\phi_B$.}
\end{figure}

In order to determine the restrictions on these parameters we have
imposed that the anti-neutrino yield should not exceed the current
experimental bound, $2.8\times 10^{-4}\phi_B$ within the presently
allowed 90\%~C.L.  region of the ($\Delta m^2_{sol}$, $\theta_{sol}$)
plane.  The results of this analysis are shown in
Fig~\ref{fig:random}.  The limits on $\mu^2_\nu b^2_{\perp_{max}}S^2$
versus $L_0$ correspond to different values of electron anti-neutrino
fluxes. The lower curve represents the current upper bound on the
product of magnetic moment and magnetic field which follows from the
recent KamLAND bound in Ref.~\cite{Eguchi:2003gg}. This is compared
with the original sensitivity expected by the KamLAND
collaboration~\cite{Busenitz:1998} (dashed line) and with the
Super-K~\cite{Gando:2002ub} bound (dot-dashed line).
One can see how Fig.~\ref{fig:random} quantitatively confirms the
expectation that the strongest limit on $\mu^2_\nu b^2_{\perp_{}}S^2$
corresponds to the case when the correlation scale $L_0$ is of the
same order as the neutrino oscillation length, $\lambda_{osc}\approx
100-200$km.

\begin{figure}[htbp]
  \centering
\includegraphics[width=.7\columnwidth]{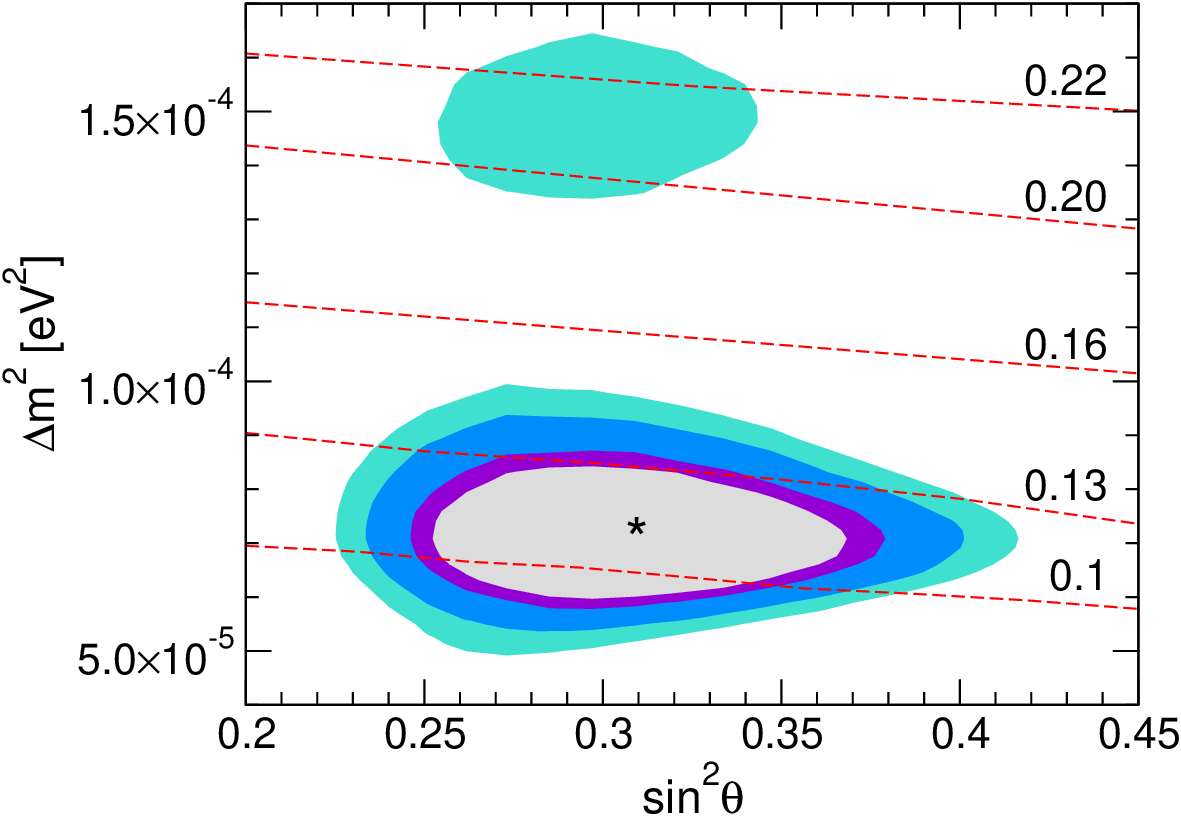}
\caption{\label{fig:iso-turb} 
  Similar as Fig.~\ref{fig:iso-random}.  The curves are iso-flux
  contours with the anti-neutrino yield fixed at the current KamLAND
  limit on $\phi_{\bar\nu_e}$ and corresponding to different values of
  $\varepsilon S \mu_{11}$. For values above 0.15 or so the high LMA
  region is ruled out. }
\end{figure}

As discussed in section~\ref{sec:magn-hydr-turb}, solar MHD turbulence
provides an attractive framework for the solar magnetic field model,
in which the anti-neutrino production probability depends only on one
extra parameter: $\mu^2_\nu\varepsilon^2 S^2$.  As we did above, we
first determine the values of $\mu^2_\nu\varepsilon^2 S^2$ that
produce an electron anti-neutrino yield of $0.028$ \% $\phi_B$.  In
Fig.~\ref{fig:iso-turb} we have shown curves corresponding to
different values of $\mu_{11} \varepsilon S $. Similarly to the case
of the simplest random field model one sees that an improved
determination of solar neutrino mixing angle will not limit $\mu_\nu
\varepsilon S $ significantly better than the current constraint. In
contrast a better determination of the solar mass splitting at KamLAND
will be useful.

\begin{figure}[htbp]
  \centering
\includegraphics[width=.7\columnwidth]{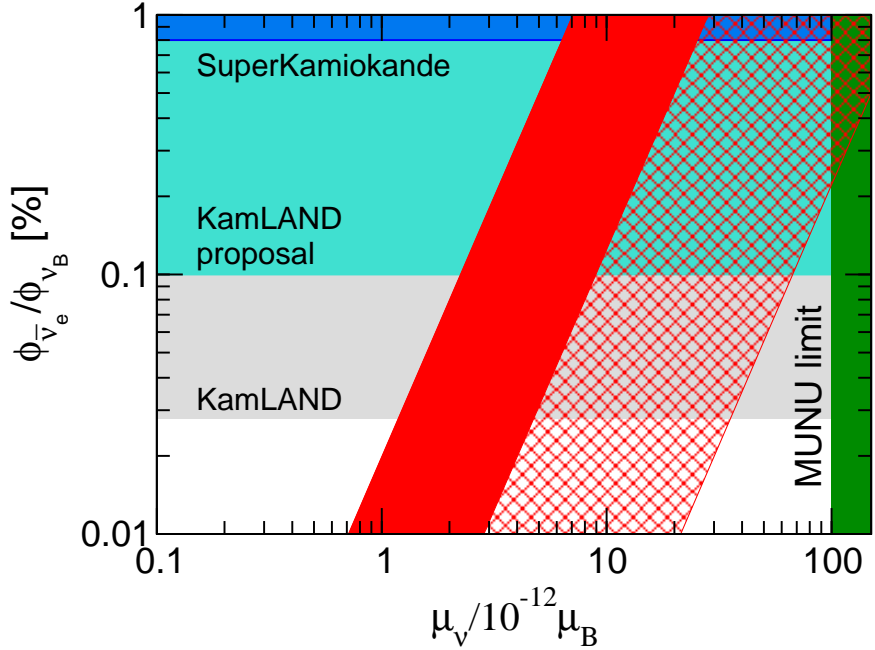}
\caption{\label{fig:turbulent} 
  Bounds on $\mu_{\nu} $ for the turbulent magnetic field model
  described in the text. The horizontal lines indicate the bounds on
  solar electron anti-neutrino fluxes from Super-K and KamLAND. The
  diagonal dark band shows the limit on the intrinsic neutrino magnetic
  moment, its width corresponds to our (Kolmogorov) turbulent magnetic
  field model uncertainties. The crossing of this band with the lowest
  KamLAND line gives the limit $\mu_\nu = 5\times10^{-12}$~$\mu_B$.
  For comparison the vertical line indicates the present MUNU reactor
  limit. A more conservative and pessimistic limit discussed in
  Sec.~\ref{sec:crit-robustn-bound} would give the limit indicated by
  the hatched band.}
\end{figure}

Following the same approach as before we have determined the limit on
$\mu_\nu \varepsilon S $ taking into account the currently 90\%~C.L.
allowed region of solar neutrino oscillation parameters. As we have
already seen in Sec.~\ref{sec:magn-hydr-turb}, a reasonable estimate
of the allowed range for $\varepsilon S$ is $0.25 < \varepsilon S <
1$. Therefore, in contrast to the previous models, with random or
regular fields, we can now, to within a factor of four, extract direct
restriction on the intrinsic neutrino magnetic transition moment
$\mu_\nu$. This is indicated in Fig.~\ref{fig:turbulent}.  The lowest
horizontal line represents current limit on the solar electron
anti-neutrino flux from the KamLAND~\cite{Eguchi:2003gg} experiment.
On the other hand our bounds on $\mu_\nu$ are given by the crossings
of the lines delimiting the dark band with the horizontal line labeled
KamLAND. 
From the Fig.~\ref{fig:turbulent} one sees that $\mu_\nu \leq
5\times10^{-12}\mu_B$~\cite{Miranda:2003yh}. For comparison the best
current laboratory limit by the MUNU experiment ($\mu_\nu < 1.0 \times
10^{-10}\mu_B$ at 90\% C.L.)~\cite{Daraktchieva:2003dr} is also
indicated. This should be compared with the best astrophysical limit,
estimated as $\mu_\nu < 3.0 \times
10^{-12}\mu_B$~\cite{Raffelt:1990pj}.
As discussed in Sec.~\ref{sec:crit-robustn-bound} under
unlikely circumstances this bound might be weakened by at most one
order of magnitude. This would correspond to the tilted line
delimiting the hatched band from the right.

An important issue arises here, namely the robustness of the bounds we
have obtained with respect to different possible choices of the
scaling law for the turbulent kinetic spectrum.
In order to answer this question we have considered the
Iroshnikov-Kraichnan model~\cite{1983flma....3.....Z}, characterized
by the power law $p=3/2$ instead of $p=5/3$ that corresponds to the
Kolmogorov spectrum. We have found that, in this case, the
anti-neutrino yield is higher by $\sim 30$\%, implying a
correspondingly stronger bound on the neutrino magnetic moment. In
general, values of $p$ lower than $5/3$ lead to the same tendency: the
smaller $p$, the stronger limit on the neutrino magnetic moment.

\begin{figure}[htbp]
  \centering
\includegraphics[width=.7\columnwidth]{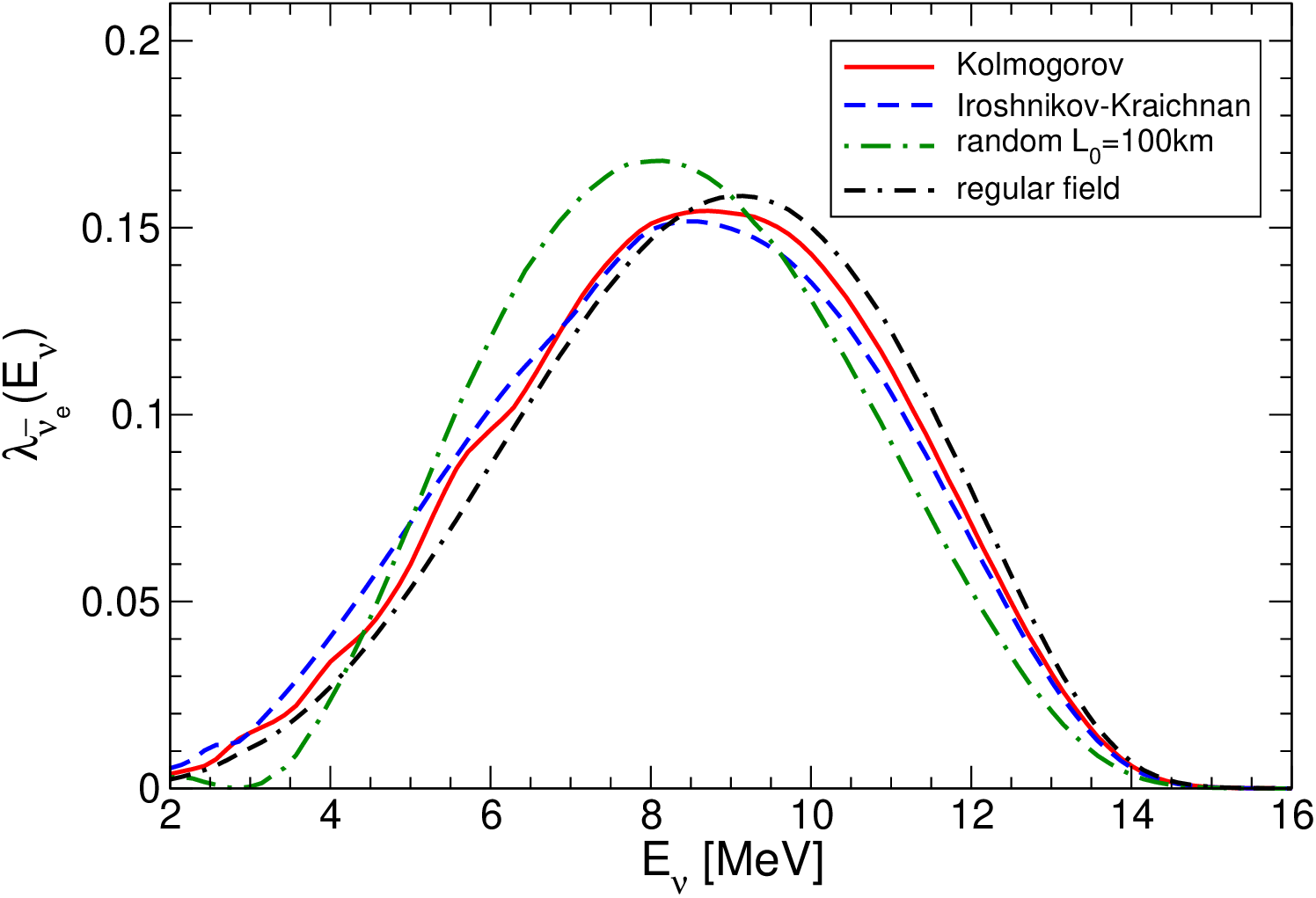}
\caption{\label{fig:spectrum} 
  Predicted normalized electron anti-neutrino spectrum for the MHD
  turbulent model with Kolmogorov (solid) or Iroshnikov-Kraichnan
  (dashed) scaling exponents and the simplest random magnetic field
  model.  For comparison we also indicate the expectation for the a
  regular magnetic field model.}
\end{figure}

Another question which may be addressed is whether different
power-laws could be distinguished in future experiments, should solar
electron anti-neutrinos ever be measured.  Although the anti-neutrino
yield is larger for the Iroshnikov-Kraichnan spectrum than for the
Kolmogorov one, the expected anti-neutrino spectrum is not
significantly different. In Fig.~\ref{fig:spectrum} the predicted
anti-neutrino spectra (normalized to unity) are plotted both for the
Iroshnikov-Kraichnan and Kolmogorov spectra. For comparison we have
also shown the expected electron anti-neutrino spectrum for the random
magnetic field case with $L_0=100$km and for the Kutvitsky-Soloviev
magnetic field scenario~\cite{Kutvitskii}. One can see that there is
no essential difference between them. Therefore, if a positive
anti-neutrino signal is ever detected, the spectrum shape would not
convey definitive information about the turbulent energy spectrum.  In
contrast, different $L_0$ values in random field scenario lead to
significantly different energy spectrum predictions. This may help in
some cases to distinguish these models, but the detailed analysis of
this phenomenon is out of the scope of the paper.

Note that in the framework of our turbulent picture there is no time
dependence in the magnetic field correlators, so that the resulting
solar neutrino fluxes will have no time variation. This is a
reasonable aproximation since the neutrino data we use are taken over
periods larger than characteristic turbulent time fluctuations of
solar magnetic fields.  Possible neutrino flux variations in the
Super-Kamiokande data have been
considered~\cite{Sturrock:2003kv,Yoo:2003rc}.
We note however that in the framework of the spin flavour precession
of Majorana neutrinos, irrespective of the magnetic field model
adopted, the expected variations in the electron and muon neutrino
fluxes should not exceed the current electron anti-neutrino flux bound
$2.8\cdot 10^{-4}\phi_B$. Therefore time variations are expected to be
too small, given the current statistical errors in the solar neutrino
data.

\section{A critique of the robustness of the bound obtained}
\label{sec:crit-robustn-bound}

We now discuss in more detail the possible weaknesses of the method we
have employed to constrain the neutrino magnetic moment in the case of
turbulent magnetic fields. The bound is given from Eq.~(\ref{Feta})
and depends on $L_{\rm{max}}$.  
We have taken
$L_{\rm{max}} \sim 1,000$ km, arguing that this is the size of the
granules on the surface of the Sun.  
One might argue, however, that the scale $L_{max}$ may be associated with
the size of the convection cells which is about a few $\times 10^4$ km.
In such a worse-case scenario the neutrino conversion would
be smaller, thus weakening the limit. However at most this would
weaken our bound by a factor $(\rm{few}\times 10)^{1/3} \sim 3$ due to
the mild power dependence 1/3. Moreover, the size of
the largest turbulent eddies, $L_{\rm{max}}$, is still an open
question, see for example Ref.~\cite{1983flma....3.....Z}.

The bound on the neutrino magnetic moment also depends on the shape
factor $ 0.5 < S <1$. Such shape factor $S$ accounts for the fact that
the random magnetic field may not be located in the whole convective
zone (see discussion after Eq.~(\ref{S})).  The bound on the neutrino
magnetic moment assuming that the shape factor is S=0.5 is a
conservative bound and corresponds to the assumption that only 1/4
part of the convective zone is effectively filled by the random
magnetic field ($S^2=0.25$).

The strength of the random (as well as the regular) magnetic field in
the solar convective zone is unknown. There are different theoretical
estimates of its value based on available experimental data. In order
to derive our conservative limit on the neutrino magnetic moment we
have used the value 50kG. A smaller value, say 20kG, for the strength
of the large--scale regular magnetic field in the convective zone
would result in a limit only 2.5 times weaker than the value we
obtained.  However we stress that the small--scale random magnetic
field is expected to be comparable or even larger than the regular
one.  The value of 50 kG for the random magnetic field does not
contradict the helioseismic data, in particular the analyses of SOHO
MDI or GONG data.  For example, the observed asphericity of the
inferred sound speed may be explained by a magnetic field at the
bottom of the convective zone of strength
200-300kG~\cite{Antia:2002ti}.  We mention also that recent results on
the modelling of the sunspot generation suggest that the strength of
the toroidal field at the base of the solar convective zone must be on
the order of 100 kG or so~\cite{newref2} in which case the bound would
get stronger.

All in all, taking all of the uncertainties mentioned above into
acount we find that, in the worst case, they will be able to weaken
our bound at most by one order of magnitude (see
Fig.~\ref{fig:turbulent}). Still the resulting bound is better than
the current direct experimental bound.  This situation would, in our
opinion, be extremely unlikely as it would require a conspiracy.

\section{Conclusions}
\label{sec:conclusions}

We have considered neutrino spin-flavor precession in stationary solar
magnetic fields, treated perturbatively with respect to LMA-MSW
neutrino oscillations.  We have discussed the impact of the recent
KamLAND constraint on the solar anti-neutrino flux on the solutions of
the solar neutrino problem in the presence of Majorana neutrino
transition magnetic moments.  This leads to strong limits on neutrino
spin-flavor precession, involving $\mu_\nu B$.  We analysed these
constraints for a number of models of solar magnetic fields, both
regular as well as random.  We found that for a turbulent solar random
magnetic field model, one can find a rather stringent constraint on
the \emph{intrinsic} neutrino magnetic moment down to the level of
$\mu_\nu \lesssim {\rm few} \times 10^{-12}\mu_B$, similar to bounds
obtained from star cooling.  Such magnetic moments would have no
effect in the detection process, given current experimental
sensitivities. We also discussed how, in the worst possible case,
these limits might deteriorate at most by one order of magnitude.
We note also that for the complementary case where there is no spin
flavor precession in the Sun and the only effect of the neutrino
magnetic moment happens in the detection process~\cite{Grimus:2002vb}
the current sensitivity is weaker than found here~\cite{Liu:2004ny}.
Therefore we conclude that turbulent solar magnetic fields provide an
enhanced sensitivity to very small neutrino transition magnetic
moments.  We have also shown how our result is rather insensitive to
the details of the assumed model of turbulent random magnetic field.
We verified this explicitly for Kolmogorov's and Iroshnikov-Kraichnan
spectra.  Should solar anti-neutrinos ever be detected, it is unlikely
that this would be of help in testing the intrinsic scaling law $B(L)
\propto L^{1/3}$ characterizing turbulence.

\section{Acknowledgements}

We thank V. B. Semikoz and D. D. Sokoloff for useful discussions.
This work was supported by Spanish grant BFM2002-00345, by European
RTN network HPRN-CT-2000-00148, by European Science Foundation network
grant N.~86, and MECD grant SB2000-0464 (TIR). TIR and AIR were
partially supported by the Presidium RAS and CSIC-RAS grants and RFBR
grant 04-02-16386.  OGM was supported by CONACyT-Mexico and SNI.


\appendix
\section{3-$\nu$ description of $\bar\nu_e$ production in solar random magnetic fields}

To justify the validity of the 2-$\nu$ scenario one can generalize it
to the 3-$\nu$ case. The probability of the solar electron
anti-neutrino appearence at the surface of the Earth (averaged over
the Sun-Earth distance) takes the form
\begin{eqnarray}
\label{Pantinu1}
P_{e\bar e}& \approx &
|\mu_{12}|^2 I_{12} \left\{ 
c_{13}^2 c_{12}^2 P_{2L}(0.7)+ c_{13}^2 s_{12}^2 P_{1L}(0.7) \right\}\nonumber\\ 
&+&|\mu_{23}|^2 I_{23} \left\{ 
c_{13}^2 s_{12}^2 P_{3L}(0.7)+ s_{13}^2 P_{2L}(0.7) \right\}\\ 
&+&|\mu_{13}|^2 I_{13} \left\{ 
c_{13}^2 c_{12}^2 P_{3L}(0.7)+ s_{13}^2 P_{1L}(0.7) \right\}\,.\nonumber 
\end{eqnarray}
Here $\mu_{jk}$ are the transition magnetic moments; $s_{jk}$ and
$c_{jk}$ are corresponding $\sin$ and $\cos$ of the 3-$\nu$ mixing
angles ($\theta_{12}$ -- solar, $\theta_{23}$ -- atmospheric,
$\theta_{13}$ -- reactor mixing angles); $P_{jL}(r=0.7R_\odot)$ are
the solar neutrino probabilities at the bottom of the convective zone.
The integrals $I_{jk}$ are the generalizations of Eq.(\ref{eta})
\begin{equation}
I_{jk} = \frac{1}{2}\int^L_0 dt_1 \int^L_0 dt_2 (b_+(t_1)b_-(t_2)
e^{-2i\delta_{jk}(t_1-t_2)} + c. c.)\,,
\label{intjk}
\end{equation}
where $\delta_{jk}=\Delta m_{jk}^2/4E$, $\Delta m_{jk}^2=m_k^2-m_j^2$.
To derive the above equation we have used the perturbative approach,
as in Sec.\ref{sec:evolution}, leaving only terms quadratic in
magnetic moments in the final probabilities.

We can notice first that, at a very good approximation, one has
$P_{3L}(0.7)=s_{13}^2\ll 1$ (see~\cite{Grimus:2002vb} and references
therein). The $\nu_{3L}$ yield in the solar core propagates as in
vacuum because matter effects are strongly suppressed, as $\delta_{23}
\approx \delta_{13} \gg V_{MSW}$.

Therefore the second and third terms in Eq. (\ref{Pantinu1}) are
proportional to $\sin^2\theta_{13}$. In order to estimate $I_{jk}$ we
consider the simple random magnetic field model discussed in
Sec.\ref{sec:simpl-rand-field}
\begin{equation}
I_{jk} = \frac{b^2 S^2}{\delta_{jk}^2}\frac{L}{L_0}\sin^2(\delta_{jk}L_0)
\label{Irandom}
\end{equation}
and the turbulent magnetic field model discussed in
Sec.~\ref{sec:magn-hydr-turb}
\begin{equation}
I_{jk} = 0.3\frac{b^2 S^2 L L_{max}}{(\delta_{jk}L_{max})^{5/3}}
\label{Iturb}
\end{equation}
Taking into account that solar and atmospheric mass splittings have
significantly different scales, $\Delta m_{12}^2 \ll \Delta m_{23}^2
\approx \Delta m_ {13}^2$, we obtain in both cases (random and
turbulent) that $I_{23} \approx I_{13} \ll I_{12}$, which means that
neutrino spin-flavor conversion in the channels $1-3$ and $2-3$ is
strongly suppressed with respect to the $1-2$ channel.

We may conclude that possible constraints on $|\mu_{23}|^2$ and
$|\mu_{13}|^2$ are very poor. They are strongly suppressed by two
facts:
\begin{itemize}
\item $s_{13}^2<0.054$ ($3\sigma$ C.L.)~\cite{Maltoni:2003da},
\item $0.018<\Delta m_{12}^2/\Delta m_{13}^2\approx\Delta
  m_{12}^2/\Delta m_{23}^2<0.053$ ($3\sigma$ C.L.)~\cite{Maltoni:2003da}.
\end{itemize}

Therefore the contribution of other channels involving $\mu_{23}$ and
$\mu_{13}$ to electron anti-neutrino production is strongly suppressed
both directly by the small value of the angle $\theta_{13}$ and by the
small ratio of solar to atmospheric squared mass differences
$\Delta_{sol}^2/\Delta_{atm}^2$. As a result we adopt the 2-$\nu$
picture, characterized by a single component of the transition
magnetic moment matrix ($\mu_{12}$) as a very good approximate
description of anti-neutrino production.



\begin{thebibliography}{10}

\bibitem{Eguchi:2003gg}
KamLAND Collaboration, K.~Eguchi {\em et~al.},
\newblock Phys. Rev. Lett. {\bf 92}, 071301 (2004), [hep-ex/0310047].

\bibitem{Gando:2002ub}
Super-Kamiokande Collaboration, Y.~Gando {\em et~al.},
\newblock Phys. Rev. Lett. {\bf 90}, 171302 (2003), [hep-ex/0212067].

\bibitem{Aharmim:2004uf}
B.~Aharmim {\it et al.}  [SNO Collaboration],
arXiv:hep-ex/0407029.

\bibitem{Schechter:1981hw}
J.~Schechter and J.~W.~F. Valle,
\newblock Phys. Rev. {\bf D24}, 1883 (1981),
\newblock Err. Phys. Rev. D25, 283 (1982).

\bibitem{akhmedov:1988uk}
E.~K. Akhmedov,
\newblock Phys. Lett. {\bf B213}, 64 (1988).

\bibitem{lim:1988tk}
C.-S. Lim and W.~J. Marciano,
\newblock Phys. Rev. {\bf D37}, 1368 (1988).

\bibitem{Eguchi:2002dm}
KamLAND Collaboration, K.~Eguchi {\em et~al.},
\newblock Phys. Rev. Lett. {\bf 90}, 021802 (2003), [hep-ex/0212021].

\bibitem{Barranco:2002te} 
  J.~Barranco {\em et~al.}, \newblock Phys.  Rev. {\bf D66}, 093009
  (2002), [hep-ph/0207326, version 3 which contains the
  KamLAND-update].

\bibitem{Ahmed:2003kj}
S.~N.~Ahmed {\it et al.}  [SNO Collaboration],
Phys.\ Rev.\ Lett.\  {\bf 92}, 181301 (2004)
[arXiv:nucl-ex/0309004].

\bibitem{Maltoni:2003da}
M.~Maltoni, T.~Schwetz, M.~A. Tortola and J.~W.~F. Valle,
\newblock Phys. Rev. {\bf D68}, 113010 (2003), [hep-ph/0309130].
For a recent review see M.~Maltoni {\em et~al.},
New J. Phys. {\bf 6} 122 (2004)
http://stacks.iop.org/1367-2630/6/122, 
arXiv:hep-ph/0405172.

\bibitem{Miranda:2003yh}
O.~G. Miranda, T.~I. Rashba, A.~I. Rez and J.~W.~F. Valle,
\newblock Phys.\ Rev.\ Lett.\  {\bf 93}, 051304 (2004) [hep-ph/0311014]

\bibitem{Bykov:1998gv}
A.~A. Bykov, V.~Y. Popov, A.~I. Rez, V.~B. Semikoz and D.~D. Sokoloff,
\newblock Phys. Rev. {\bf D59}, 063001 (1999), [hep-ph/9808342].

\bibitem{Kutvitskii}
V.~A. Kutvitskii and L.~S. Solov'ev,
\newblock J. Exp. Theor. Phys. {\bf 78}, 456 (1994).

\bibitem{Guzzo:1998sb}
M.~M. Guzzo and H.~Nunokawa,
\newblock Astropart. Phys. {\bf 12}, 87 (1999), [hep-ph/9810408].

\bibitem{akhmedov:2002mf}
E.~K. Akhmedov and J.~Pulido,
\newblock Phys. Lett. {\bf B553}, 7 (2003), [hep-ph/0209192].
B.~C.~Chauhan, J.~Pulido and E.~Torrente-Lujan,
\newblock Phys.\ Rev.\ D {\bf 68} (2003) 033015
[arXiv:hep-ph/0304297].

\bibitem{Friedland:2002pg}
A.~Friedland and A.~Gruzinov,
\newblock Astropart. Phys. {\bf 19}, 575 (2003), [hep-ph/0202095].

\bibitem{1983flma....3.....Z}
I.~B. {Zeldovich}, A.~A. {Ruzmaikin} and D.~D. {Sokolov},
\newblock {\em {Magnetic fields in astrophysics}} (New York, Gordon and Breach
  Science Publishers), 1983.

\bibitem{Raffelt:1990pj}
G.~G. Raffelt,
\newblock Phys. Rev. Lett. {\bf 64}, 2856 (1990).

\bibitem{Torrente-Lujan:2003cx}
E.~Torrente-Lujan,
\newblock JHEP {\bf 04}, 054 (2003), [hep-ph/0302082].

\bibitem{Loreti:1994ry}
F.~N. Loreti and A.~B. Balantekin,
\newblock Phys. Rev. {\bf D50}, 4762 (1994), [nucl-th/9406003].

\bibitem{Nunokawa:1996qu}
H.~Nunokawa, A.~Rossi, V.~B. Semikoz and J.~W.~F. Valle,
\newblock Nucl. Phys. {\bf B472}, 495 (1996), [hep-ph/9602307].

\bibitem{Balantekin:2003qm}
A.~B. Balantekin and H.~Yuksel,
\newblock Phys. Rev. {\bf D68}, 013006 (2003), [hep-ph/0303169].

\bibitem{Guzzo:2003xk}
M.~M. Guzzo, P.~C. de~Holanda and N.~Reggiani,
\newblock Phys. Lett. {\bf B569}, 45 (2003), [hep-ph/0303203].

\bibitem{Nicolaidis:1991da}
A.~Nicolaidis,
\newblock Phys. Lett. {\bf B262}, 303 (1991).

\bibitem{Semikoz:1998ef}
V.~B. Semikoz and E.~Torrente-Lujan,
\newblock Nucl. Phys. {\bf B556}, 353 (1999), [hep-ph/9809376].

\bibitem{Burgess:1997mz}
C.~P. Burgess and D.~Michaud,
\newblock Annals Phys. {\bf 256}, 1 (1997), [hep-ph/9606295].

\bibitem{Burgess:2002we}
C.~Burgess {\em et~al.},
\newblock Astrophys. J. {\bf 588}, L65 (2003), [hep-ph/0209094].

\bibitem{Burgess:2003su}
C.~P. Burgess {\em et~al.},
\newblock JCAP {\bf 0401}, 007 (2004), [hep-ph/0310366].

\bibitem{Bamert:1998jj}
P.~Bamert, C.~P. Burgess and D.~Michaud,
\newblock Nucl. Phys. {\bf B513}, 319 (1998), [hep-ph/9707542].

\bibitem{Subramanian:1999mr}
K.~Subramanian,
\newblock Phys. Rev. Lett. {\bf 83}, 2957 (1999), [astro-ph/9908280].

\bibitem{rez04}
O.~G. Miranda {\em et~al.},
\newblock Proc. of International Workshop on Astroparticle and High Energy
  Physics, October 14 - 18, 2003, Valencia, Spain, published at JHEP,
  PRHEP-AHEP2003/072, accessible from http://ific.uv.es/ahep/.

\bibitem{Miranda:2000bi}
O.~G. Miranda {\em et~al.},
\newblock Nucl. Phys. {\bf B595}, 360 (2001), [hep-ph/0005259].

\bibitem{MoninYaglom}
A.~S. Monin and A.~M. Yaglom,
\newblock Statistical Fluid Mechanics, MIT Press, (Cambridge, 1975).

\bibitem{Kolmogorov:1941}
A.~N. Kolmogorov,
\newblock Dokl. Akad. Nauk. SSSR. {\bf 31}, 538 (1941).

\bibitem{Kolmogorov:1991}
A.~N. Kolmogorov,
\newblock Proc. Roy. Soc. Lond. {\bf A434}, 15 (1991).

\bibitem{Goldreich}
P.~{Goldreich} and S.~{Sridhar},
\newblock Astrophys. J. {\bf 438}, 763 (1995).

\bibitem{Cho}
J.~{Cho} and E.~T. {Vishniac},
\newblock Astrophys. J. {\bf 539}, 273 (2000).

\bibitem{1991sun..book.....S}
M.~{Stix},
\newblock {\em {The Sun. an Introduction}} (Springer-Verlag Berlin Heidelberg
  New York.

\bibitem{Bahcall:2004fg}
J.~N.~Bahcall and M.~H.~Pinsonneault,
Phys.\ Rev.\ Lett.\  {\bf 92}, 121301 (2004)
[arXiv:astro-ph/0402114].

\bibitem{Busenitz:1998}
KamLAND Collaboration, J.~Busenitz {\em et~al.},
\newblock Stanford-HEP-98-03, [http://kamland.lbl.gov/TalksPaper/].

\bibitem{Daraktchieva:2003dr}
MUNU Collaboration, Z.~Daraktchieva {\em et~al.},
\newblock Phys. Lett. {\bf B564}, 190 (2003), [hep-ex/0304011].

\bibitem{Sturrock:2003kv}
P.~A. Sturrock,
\newblock Astrophys. J. {\bf 605}, 568 (2004), [hep-ph/0309239].

\bibitem{Yoo:2003rc}
Super-Kamiokande Collaboration, J.~Yoo {\em et~al.},
\newblock Phys. Rev. {\bf D68}, 092002 (2003), [hep-ex/0307070].

\bibitem{Grimus:2002vb}
W.~Grimus {\em et~al.},
\newblock Nucl. Phys. {\bf B648}, 376 (2003), [hep-ph/0208132].

\bibitem{Antia:2002ti}
H.~M.~Antia, S.~M.~Chitre and M.~J.~Thompson,
Zone,''
Astron.\ Astrophys.\  {\bf 399}, 329 (2003)
[arXiv:astro-ph/0212095].

\bibitem{newref2}
D. Nandy and A. Rai Choudhuri, 
\newblock  Science 296 (2002) 1671.

\bibitem{Liu:2004ny}
D.~W.~Liu {\it et al.}  [Super-Kamiokande Collaboration],
\newblock arXiv:hep-ex/0402015.

\end{thebibliography}
\end{document}